# Hashing and metric learning for charged particle tracking

**Sabrina Amrouche**[*]  **Moritz Kiehn**  **Tobias Golling**
Université de Genève

**Andreas Salzburger**
European Organization for Nuclear Research CERN

## Abstract

We propose a novel approach to charged particle tracking at high intensity particle colliders based on Approximate Nearest Neighbors search. With hundreds of thousands of measurements per collision to be reconstructed e.g. at the High Luminosity Large Hadron Collider, the currently employed combinatorial track finding approaches become inadequate. Here, we use hashing techniques to separate measurements into buckets of 20-50 hits and increase their purity using metric learning. Two different approaches are studied to further resolve tracks inside buckets: Local Fisher Discriminant Analysis and Neural Networks for triplet similarity learning. We demonstrate the proposed approach on simulated collisions and show significant speed improvement with bucket tracking efficiency of 96 % and a fake rate of 8 % on unseen particle events.

## 1 Introduction

Modern high energy hadron collider experiments, such as the ATLAS and CMS experiments at the LHC at CERN, produce vast amounts of data to be analyzed. This is driven by the fact that in order to maximize the research potential, many proton-proton collisions are performed with many particles emerging from the interaction region. Future upgrades of existing or new accelerators such as the High Luminosity Large Hadron Collider by 2026 or the proposed FCC collider will significantly increase this complexity. Finding the trajectories of particles produced in such collisions is a crucial step and yet, due to limitations of the current approaches, it is the most CPU intensive task in reconstruction. A trajectory is the trail a particle leaves when interacting with the detector material and causing a detectable signal in the traversed devices. The measurements are referred to as hits and the process of grouping together the hits generated by the same particle is called tracking.

Tracking in high energy physics experiments is particularly challenging due to the high number of trajectories that need to be resolved simultaneously(1). This is mainly due to the absence of hit-particle descriptive characteristics (identity) hence making them all possible candidates for a trajectory. Any two points can potentially be associated into a trajectory that can in turn be evaluated only after it is completed. This would be comparable to tracking cars that change colors and shapes at every frame[2]. In this study we consider high-luminosity scenarios where the target is to rapidly label more than 100K hits generated by more than 10K particles. We propose to combine search techniques and deep learning to solve efficiently the tracking problem.

---

[*]Corresponding author

[2]In the contrast to cars, however, we have quasi-deterministic equations of motions that govern the particle trajectory.

## 2 Proposed approach: hashing and metric learning

Finding similar items in large datasets is a challenging yet well studied problem. A high dimensionality of the records makes it even more problematic to define a valid similarity metric. Approximate Nearest Neighbors (ANN) techniques are often used to reduce the search complexity(2). Formally the ANN problem is defined as follows: Given a set $S$ of $n$ data points in a metric space $X$ the task is to index the points so that, given a query point $q \in X$, the data points most *similar* to $q$ are quickly found. ANN indexes items into *hashes or buckets* that contain approximate neighbors to a query point. As a result, for a given item of interest, searching for its *neighbors* becomes restricted to the bucket only.

We propose to use ANN to index particle tracks and therefore reduce the tracking to only be applied within a single bucket. The neighbors of a given query point are the hits from its true particle. The main appeal of ANN for tracking is the fast index query time that allows to potentially retrieve a particle from one query point. Moreover, by using queries across the hit point cloud, the approach is easily parallelizable. The ANN notion of similarity is redefined to find neighbors as points aligned along the same trajectory (particle neighbors) by using a suitable distance function. The hashes can be regarded as *average* track estimates that are refined with metric learning. Within a bucket, we use metric learning to map to an optimal embedding where hits from the same particle naturally cluster with a simple metric.

For the conducted experiments we use the TrackML challenge dataset(3). This dataset simulation is done assuming an average of 200 overlaid simulatenous (so called pile-up) events with one signal event. The latter has been chosen to be a top quark pair production, whereas for the scope of this study only minor importance is given to the actual physics event. The data points are simulated by ACTS[3], a fast simulation emulating particle collisions as they might occur in a detector with high accuracy. A particle leaves on average 12 hits in the simulated TrackML detector where each hit is characterized by 5 features: global coordinates $x, y, z$ and two angles $\phi$ and $\theta$ that are obtained from the cluster shape and indicate the track direction with respect to the detection sensor. In the remainder of this paper we describe the two components of the proposed approach applied to the TrackML dataset.

### 2.1 Reducing complexity with hashing

We choose to build the ANN index as a tree based partition method (binary trees) where the points are assigned according to their feature values only (unsupervised). In the first experiment, we randomly query hits and request their closest neighbors. We analyze then the content of each bucket and record the size of the leading (largest) track. The average query time is $0.05\,\mathrm{ms}$ on an Intel(R) Core(TM) i7-6500U CPU @ 2.50GHz.
Figure 1 shows the size of the leading particle in a bucket of 20 neighbors and 50 neighbors for 5000 random queries. The ANN index was built on $[x, y, z]$ vectors with the angular distance using the open source ANN library Annoy (4) and without any pre-filtering of the dataset (100K hits). The particle size distribution shows that most buckets contain valid particles with up to 15 hits from the same particle in a 20 neighbors bucket.

Since the particle label associated with every point is available from simulation, we can measure the number of queries necessary to reconstruct a full event assuming an optimal bucket tracking[4]. A track will be marked as reconstructed if $80\,\%$ of its hits are found inside the same bucket (purity $\geq 0.8$). Figure 2 shows the efficiencies for bucket sizes of 20 and 50 as a function of the number of queries. The efficiency is defined as the number of reconstructed tracks divided by the total number of tracks. In Figure 2 we consider only particles with a transverse momentum $p_T$ greater than $1\,\mathrm{GeV}$ and at least 4 hits per track. With 2000 random queries only, 20 neighbors buckets retrieve slightly more than $95\,\%$ particles while 50 neighbors buckets retrieve over $99.99\,\%$. The efficiencies are for $\mu = 200$ events. Querying sequentially 2000 random buckets of 50 hits takes $0.09\,\mathrm{s}$ on the architecture previously mentioned and with the python API provided by the Annoy library.

---

[3] A Common Tracking Software open source library on `cern.ch/acts`
[4] Indeed, current pattern recognition algorithms at the LHC show a close to 100% efficiency for reconstructable particles, i.e. particles that leave sufficient hits in the detector to be found

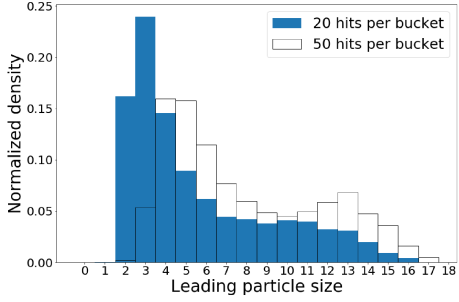 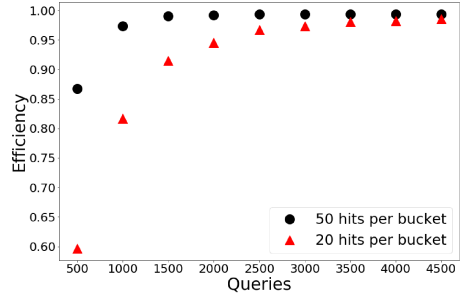

Figure 1: Leading particle size distribution   Figure 2: Efficiency using truth tracking in buckets

## 2.2 Learning a mapping for tracks

The hit features created by the particle may likely not be the optimal parameter space to eventually separate hits from same particles in close-by neighbour regions. For this reason, we propose metric learning(5)(6) to project our input dataset into a space where a more optimal clustering can be achieved. We investigate two types of metric learning approaches: Local Fisher Discriminant Analysis (LFDA) and Neural Networks.

### 2.2.1 Local Fisher Discriminant Analysis

LFDA works by maximizing between-class separability and preserving the within-class local structure at the same time(5). The latter being the particle trajectory in our study. Different metric learning techniques have been tested such as Information Theoretic Metric Learning (ITML)(6) and Large Margin Nearest Neighbor (LMNN) but LFDA gave the best results in our specific application.

The LFDA learns a projection space $[u, v]$ using the particle labels. Figure 3 shows the projection of full tracks into a learnt features space $[u, v]$ with LFDA using as input features $[\sqrt{x^2 + y^2}, \theta, \phi]$. Every segment on the figure represents a full track of at least 5 hits and a $p_T$ above 2 GeV in the original space. The segments are colored with their true associated pseudorapidity (eta) a spatial coordinate describing the angle of a particle relative to the beam axis. The correlation to eta values demonstrates the strong impact of the incidence angle $\theta$ used to learn the mapping. Considering this mapping, the task of finding tracks becomes trivial (at least if they are not inside jets) and we propose to solve it with a simple clustering algorithm. Running a KMeans with an euclidean distance on 2000 buckets, we are able to retrieve 95 % particles with buckets of 50 hits and with a $p_T$ cut of 1 GeV. These results have been achieved on an unseen dataset and similar to section 2.1, we consider a track purity of 80 %.

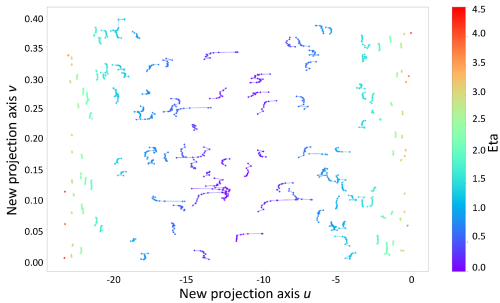 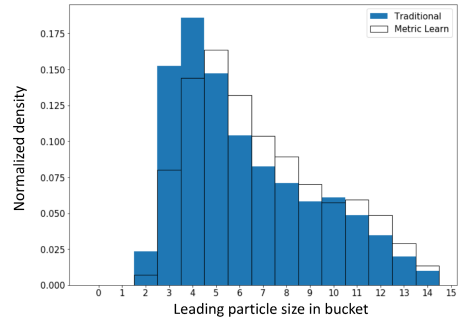

Figure 3: New tracking space with LFDA.   Figure 4: Metric learning on hashing.

We studied the impact of LFDA metric learning prior to the hashing phase. Projecting the dataset with LFDA is indeed found to increase the ANN hashes purity. Figure 4 shows the leading particle size in 20 neighbors buckets for raw $x, y, z$ (as shown in Figure 1) in contrast to the leading particle size for learnt projections using LFDA.

### 2.2.2 Neural Networks for learning triplet similarity

In this section we propose to learn a *triplet-wise* metric with a deep network. We choose to focus on the area of the detector closest to the interaction point, i.e. pixel volume. The network takes three vectors as input and computes a distance. We propose to learn the distance between triplets such as the network takes three hits and learns their similarity. We choose triplets instead of doublets because of the shape information encoded in three points. Particles have approximately helix trajectories and curvature estimation requires three points.

Data points from the same particle are labeled, using truth information from simulation, with a distance of 0 and measurements from different particles are labelled with a distance of 1. Fake triplets are randomly sampled within a bucket where the hits belong to three different layers in order to focus on difficult triplets and improve the network accuracy. Figure 5 illustrates an example bucket in an $r - z$ (longitudinal) plane of the specific detector region. Only a subset of possible combinations is shown in the figure. The colors in Figure 5 encode predictions from a trained network. We can see the confidence shown by the network on different triplets. Good triplets are found with large confidence, i.e. distance prediction close to 0 and fake triplets are marked fake with large confidence as well despite being not trivial in shape.

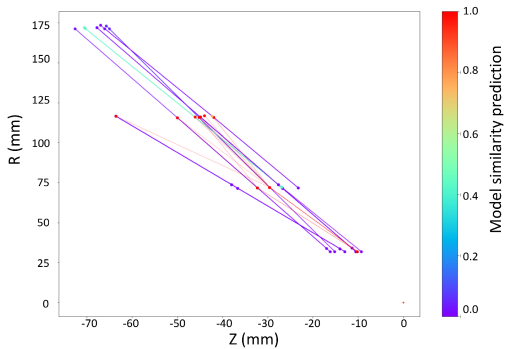
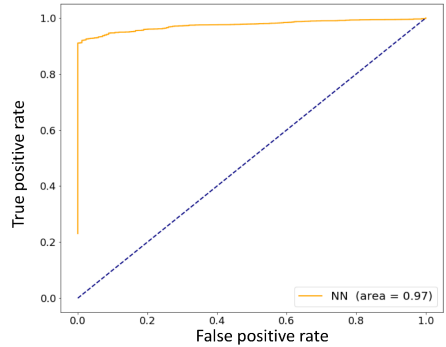

Figure 5: Triplet similarity prediction    Figure 6: ROC curve

As input to the network, each hit is represented by its global coordinates $x, y, z$ and pairwise distances of the incidence angles $\delta\theta_{ij}$ and $\delta\phi_{ij}$. Scaling the dataset is generally found to improve the learning accuracy. In our use case, we found that scaling the data with a robust scaler that removes the median and scales according to the quantile range contributed an improvement of $15\,\%$ to the area under curve AUC compared to using normalization (scaling to vector length). The network architecture is 4 dense layers with an Adagrad optimizer and a binary cross entropy loss. The triplets predicted to be from the same particle are grouped with connected component (CC) algorithm on the basis of a unique hit identification. The neural net is trained and validated on samples from pixel volume from different events. Figure 6 shows the ROC curve of the model on unseen events with excellent sensitivity and specificity (AUC=$97\,\%$). In particle tracking, finding all good triplets is as important as avoiding false positives. After applying CC on predicted triplets and considering a track purity of $60\,\%$, the overall tracking efficiency is of $86\,\%$ from the total volume pixel tracks above $1\,\text{GeV}$ and with at least 5 hits. The tracking inefficiency is exclusively attributed to inefficiency of the neural network since CC only connects triplets that are predicted to be from the same particle. False positives are therefore wrongly appended to the found tracks and false negatives are lost tracks (inefficiency).

## 3  Conclusion

In this paper we presented a new particle tracking approach consisting of Approximate Nearest Neighbors search reducing the tracking complexity to buckets followed by metric learning inside buckets to reconstruct tracks. The reduction of tracking complexity is significant as compared to current combinatorial techniques that consider several hundreds hits at every step. Using hashing also allows a constant and fast query time in the order of $0.05\,\text{ms}$ in our study setup. The speed has the potential to scale almost linearly when running on multiple processes. We also proposed an alternative to tracking in buckets: LFDA and Neural Nets which both showed promising accuracy results in finding tracks with up to $95\,\%$ tracking efficiency. A future research direction is to extend the similarity learning to quadruplets and ultimately to the full bucket.